\documentclass[%
 reprint,
 superscriptaddress,
 amsmath,amssymb,
 aps,
]{revtex4-2}

\usepackage{graphicx}% Include figure files
\usepackage{dcolumn}% Align table columns on decimal point
\usepackage{bm}% bold math
\usepackage{physics}
\usepackage{xcolor}
\usepackage[normalem]{ulem} %%%

\usepackage{pdfpages} % include pdfs
\usepackage{pgffor} % for loops

\makeatletter
\AtBeginDocument{\let\LS@rot\@undefined}
\makeatother

\def\supplementfilename{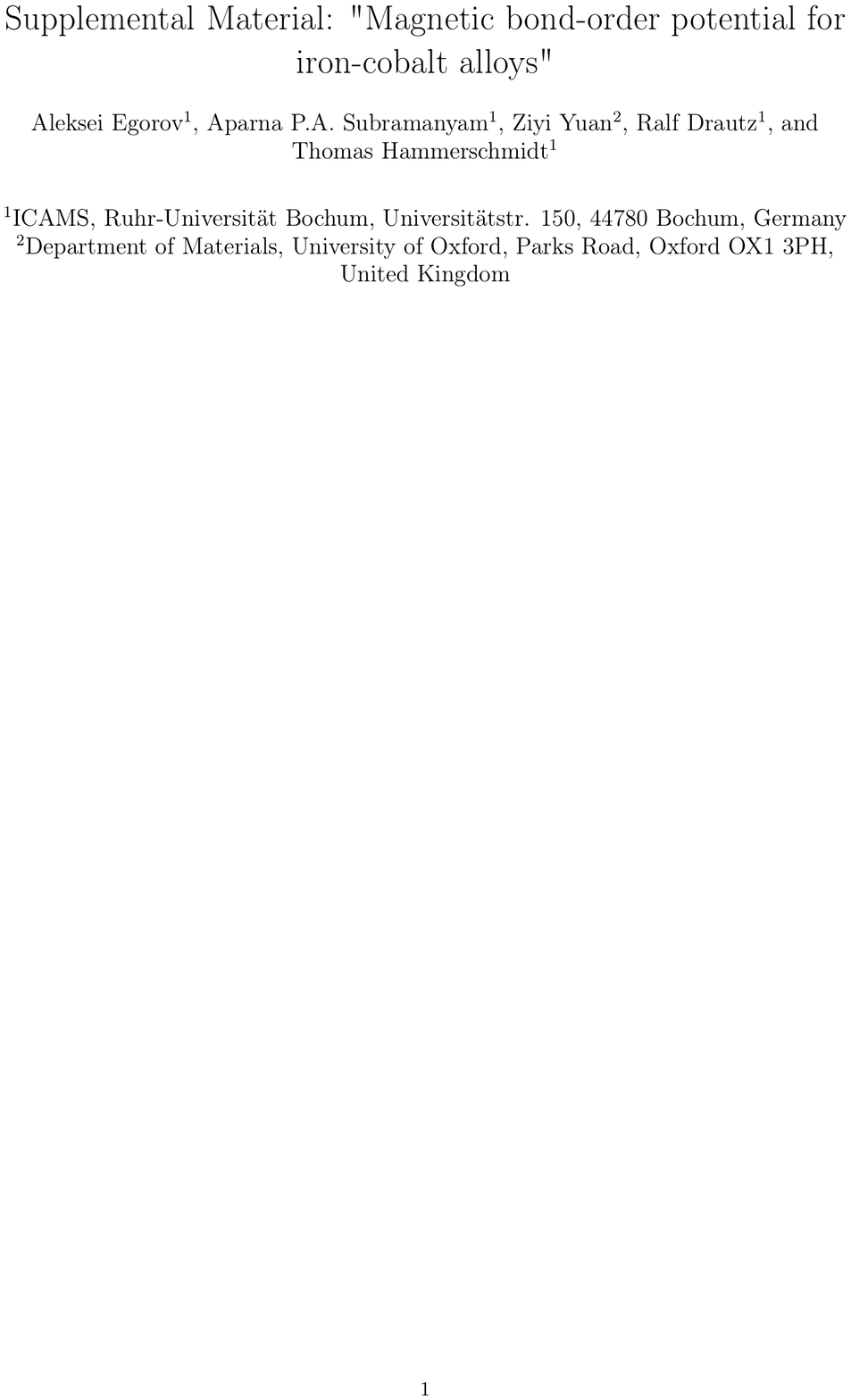}

\pdfximage{\supplementfilename}
\def\numbersupplementpages{\the\pdflastximagepages}

\newif\ifarXiv
\arXivtrue

\newcommand{\RNum}[1]{\uppercase\expandafter{\romannumeral #1\relax}}

\begin{document}

\preprint{APS/123-QED}

\title{Magnetic bond-order potential for iron-cobalt alloys}

\author{Aleksei~Egorov}
\affiliation{ ICAMS, Ruhr-Universit\"{a}t Bochum, Universit\"{a}tstr. 150, 44780 Bochum, Germany }

\author{Aparna~P.A.~Subramanyam}
\affiliation{ ICAMS, Ruhr-Universit\"{a}t Bochum, Universit\"{a}tstr. 150, 44780 Bochum, Germany }

\author{Ziyi~Yuan}
\affiliation{ Department of Materials, University of Oxford, Parks Road, OX1 3PH, United Kingdom }

\author{Ralf~Drautz}
\affiliation{ ICAMS, Ruhr-Universit\"{a}t Bochum, Universit\"{a}tstr. 150, 44780 Bochum, Germany }

\author{Thomas~Hammerschmidt}
\affiliation{ ICAMS, Ruhr-Universit\"{a}t Bochum, Universit\"{a}tstr. 150, 44780 Bochum, Germany }

\date{\today}

\begin{abstract}
For large-scale atomistic simulations of magnetic materials, the interplay of atomic and magnetic degrees of freedom needs to be described with high computational efficiency. Here we present an analytic bond-order potential (BOP) for iron-cobalt, an interatomic potential based on a coarse-grained description of the electronic structure. We fitted BOP parameters to magnetic and non-magnetic density-functional theory (DFT) calculations of Fe, Co, and Fe-Co bulk phases. Our BOP captures the electronic structure of magnetic and non-magnetic Fe-Co phases. It provides accurate predictions of structural stability, elastic constants, phonons, point and planar defects, and structural transformations. It also reproduces the DFT-predicted sequence of stable ordered phases peculiar to Fe-Co and the stabilization of B2 against disordered phases by magnetism. Our Fe-Co BOP is suitable for atomistic simulations with thousands and millions of atoms.\end{abstract}

\keywords{FeCo, Fe-Co, bond-order potential, structural stability, defects}
\maketitle

\section{\label{sec:Introduction}Introduction}

Fe-Co is a magnetic alloy with the highest saturation magnetization among all materials~\mbox{\cite{Weiss1912,Weiss1929}}, which places it at the top of the Slater-Pauling curve~\mbox{\cite{Slater1936,Pauling1938}}. Together with high permeability and high Curie temperature, Fe-Co alloys are suitable for high-performance transformers, solenoid valves, magnetostrictive transducers, and others~\cite{SunDeevi05}.

Fe-Co alloys tend to order, so the B2 (CsCl) phase, where Co atoms are at the corners of the bcc conventional unit cell and Fe at the center, forms below 1000~K at 1:1 composition~\cite{ExpB2bccTrans,ExpB2bccTrans2}. Above this temperature, the alloy transforms to disordered bcc. Abrikosov~$et$~$al.$~\cite{AbrikFeCoPRB}, using DFT, revealed that at 0~K, ordered and partially ordered ferromagnetic B2 are the most stable phase from 0 to 85 at.~$\%$ of Co. 
Neumayer and Fähnle~\cite{FahnleAntisite} addressed the role of magnetism for the ordering in light of structural defects. They found that Fe and Co antisite defects in the B2 phase have small but positive formation energies for magnetic DFT calculations. Positive formation energy prevents the spontaneous exchange of Fe and Co atoms and transformation to disordered bcc. For the non-magnetic case, antisite defects also have small but negative formation energies, making B2 unstable compared to disordered bcc. Thus, magnetism is crucial for structural stability in Fe-Co.

Large-scale atomistic simulations of magnetic alloys such as Fe-Co are still challenging. Modeling and simulation of lattice defects like dislocations, grain boundaries, or cracks often require cell sizes from a few thousand to hundreds of thousands of atoms~\cite{cellsizedisl-1,cellsizeGB-1,cellsizeGB-2,cellsizeGB-3,maresca2018screw, kermodecrack2014}. DFT provides high accuracy but is limited to only a few hundred atoms. Simulations with many atoms are possible with empirical interatomic potentials, but they usually do not capture the electronic structure and magnetism.
For this reason, for example, empirical potentials are unreliable in simulations of dislocations in bcc transition metals~\cite{DislFeCo,DudarEmpPotFailue} and, consequently, their alloys, including Fe-Co. Machine learning potentials combine high precision and computational efficiency but require a vast amount of reference DFT data~\cite{deringer2019MLreview-1,mishin2021MLreview-2,kulik2022MLreview-3,freitas2022mlp}. Besides,  incorporating magnetism is still tricky, and the first magnetic machine learning potentials are just appearing~\cite{eckhoff2021magMLp,drautz2020magace, novikov2022magMLp}. Thus, accurate electronic structure methods with lower computational costs than DFT are necessary for simulations of magnetic transition-metal alloys.

Analytic bond order potentials (BOPs) are an electronic structure method derived from quantum mechanics with an explicit treatment of magnetism~\cite{DraPett06,DraPettMagBOP11}. Analytic BOPs and the underlying tight-binding bond model have shown their applicability to non-magnetic transition metals, such as Ta, W, Nb, Mo~\cite{MirMoTaNbW14}, and Ti~\cite{TiAlbFer19}, as well as the magnetic transition metals Mn~\cite{MnMagBOPDrain14} and Fe~\cite{MroFePRL11,MadsenFe11,Ford-14}. BOP was also successfully used to study iron's magnetic~\cite{Wang-2019} and structural~\cite{Wang-2022} phase transitions at finite temperatures. At the same time, analytic BOPs are efficient enough for simulations with millions of atoms~\cite{TEIJEIROmillion1,TEIJEIROmillion2}. 

In this work, we present an analytic BOP for Fe-Co alloys. We discuss in detail the parameterization procedure, including reference data, fitting process, and validation strategies. We further validate the BOP by comparing central material properties such as energy-volume curves, phonons, the electronic density of states, defect formation energies, and elastic constants with DFT and experimental data. We further compute the convex hull of Fe-Co across the whole composition range and find that it contains a dense sequence of ordered, stable phases in excellent agreement with DFT calculations.

\section{\label{sec:abop}Analytic bond-order potentials}

Analytic bond-order potentials are interatomic potentials derived from quantum mechanics by a two-step coarse-graining from DFT to a tight-binding~(TB) bond model~\cite{Sutton-88} and from TB to BOPs~\cite{DraPettMagBOP11,Drautz-15}. 
The BOPs transparently describe the interatomic interactions and scale linearly with the system size~\cite{bopfox}. As a result, BOPs allow simulations of systems with millions of atoms~\cite{TEIJEIROmillion1,TEIJEIROmillion2}. 

The total binding energy of the BOP is given as 
\begin{equation}\label{eq:Etot}
E_\mathrm{tot} = E_\mathrm{bond} + E_\mathrm{emb} + E_\mathrm{mag} + E_\mathrm{rep},
\end{equation}
with the bond energy $E_\mathrm{bond}$, the embedding energy $E_\mathrm{emb}$,  the magnetic energy $E_\mathrm{mag}$, and the repulsive energy $E_\mathrm{rep}$, just as in the TB bond model.

The bond energy $E_\mathrm{bond}$ results from the formation of chemical bonds of the atom-centered orbitals of neighboring atoms. It is determined by integrating the local density of states (DOS) $n_{I\alpha,s}$ up to the Fermi energy $E_\mathrm{F}$,
\begin{equation} \label{eq:calcEbond}
 E_\mathrm{bond} = \sum_{I\alpha,s}\int^{E_\mathrm{F}} (E-E_{I\alpha})n_{I\alpha,s}(E)dE,
\end{equation}
of orbitals $\alpha$ of atom $I$ with onsite levels $E_{I\alpha}$ and $s$ is either spin-up or spin-down.
The onsite levels $E_{I\alpha}$ are obtained in a self-consistency loop that minimizes the total energy $E_\mathrm{tot}$ with respect to $E_{I\alpha}$ (see Ref.~\cite{bopfox} for details).
In TB, the DOS is computed by diagonalization of the system-wide Hamiltonian $\hat{H}$, while in analytic BOP, from the moments of the local DOS
\begin{equation} \label{eq:calc of moments}
 \mu_{I\alpha}^{(p)} = \int E\strut^{p} n_{I\alpha}(E)dE,
\end{equation}
\noindent for the non-magnetic case, where $\mu_{I\alpha}^{(p)}$ is the $p$-th moment of the DOS of orbital $\alpha$ of atom $I$. We can also compute moments from all self-returning hopping paths of length $p$ that start and end at $\ket{I\alpha}$ as
\begin{equation}\label{eq:hopping paths}
\begin{split}
\mu^{(p)}_{I\alpha} & = \bra{I\alpha}\hat{H}\strut^p \ket{I\alpha} \\
& = \sum\limits_{J\beta K\gamma...} \bra{I\alpha}\hat{H} \ket{J\beta}\bra{J\beta}\hat{H} \ket{K\gamma}\bra{K\gamma}\hat{H} \ket{...} \\ & \times ...\bra{...}\hat{H} \ket{I\alpha} \\
& = \sum\limits_{J\beta K\gamma...} H_{I\alpha J\beta} H_{J\beta K\gamma} H_{K\gamma...}...H_{...I\alpha},
\end{split}
\end{equation}
with the Hamiltonian matrix elements $H_{I\alpha J\beta}$ between pairs of orbitals $\alpha$ and $\beta$ on atoms $I$ and $J$. For the magnetic case, we need to evaluate moments in Eq.~\ref{eq:calc of moments} and Eq.~\ref{eq:hopping paths} separately for spin-up and spin-down and then sum them up (see Ref.~\cite{DraPettMagBOP11} for details).
Non-diagonal matrix elements $H_{I\alpha J\beta}$ in Eq.~\ref{eq:hopping paths}, also known as hopping or bond integrals, are distance-dependent. We can represent them in the parameterized form (Eq.~\ref{eq:exponents}) to later fit parameters to reference data, as discussed in section~\ref{sec:Construction}.
The equality of Eq.~\ref{eq:calc of moments} and Eq.~\ref{eq:hopping paths} directly relates the electronic structure (DOS) to the atomic structure (self-returning hopping paths). It is a core feature of the BOP methodology~\cite{CyrotLackMom}.

Higher moments (with a higher value of $p$ in Eq.~\ref{eq:calc of moments}) correspond to longer hopping paths and hence a more far-sighted exploration of the local atomic environment.
Four moments are sufficient to distinguish fcc and bcc structures in transition metals~\cite{DraHammBOP09chap}. Six moments are needed to capture the trend of structural stability hcp$\rightarrow$bcc$\rightarrow$hcp$\rightarrow$fcc across the 4d and 5d transition metal series~\cite{DraPett06}. In the limit of an infinite number of moments, we would recover the TB solution. 
For our Fe-Co BOP, we used nine moments as in an existing BOP for iron~\cite{Ford-14} that showed that nine moments are an appropriate trade-off between computational efficiency and accuracy.
We employed the Jackson kernel with 200 extended moments to smoothen the truncation of the BOP expansion with a constant terminator as described in Refs.~\cite{bopfox,seiser2013terminator}. The analytic BOP calculations in this work are performed with the BOPfox \mbox{software package~\cite{bopfox}}.

The setup of the pairwise Hamiltonian $H_{I\alpha J\beta}$ for TB/BOP calculations depends on the system. Both $s$ and $d$ electrons in Fe and Co contribute to bond formation. 
We approximate the interatomic interaction with an orthogonal, $d$-valent Hamiltonian in a two-center approximation. Cawkwell $et$ $al.$ showed that neglecting $sp$ and $sd$ interactions is an excellent approximation for iridium~\cite{Cawkwell-06}; therefore, it is justified for Fe-Co with similar band-filling.
The resulting 5$\times$5 Hamiltonian of two interacting $d$-valent atoms $I$ and $J$ at a distance $R$ contains three independent bond integrals $dd\sigma$, $dd\pi$, and $dd\delta$ that we represent as a sum of exponential functions
\begin{equation}
\beta(R) = \sum\limits_{k=1}^{k_\mathrm{max}} c_k \exp (-\lambda_k R^{n_k}),
\label{eq:exponents}
\end{equation}
where $c_k$, $\lambda_k$, and $n_k$  are the parameters to fit. This simple functional form has been used successfully for the bond integrals obtained from projecting the DFT eigenstates onto an orthogonal TB basis~\cite{JenkeDimers}. For Fe-Co BOP, we found that three exponentials ($k_\mathrm{max}=3$) for the $dd\sigma$ bond and two ($k_\mathrm{max}=2$) for $dd\pi$ and $dd\delta$ provide a numerically robust optimization and sufficient flexibility to reach the target precision.

Omitting the $s$ electrons is compensated by the embedding energy $E_\mathrm{emb}$ in Eq.~\ref{eq:Etot}.
Following the second-moment approximation of TB, where the cohesive energy of an atom $i$ scales as the square root of the densities of the surrounding atoms, we express the embedding energy as
\begin{equation}
E_\mathrm{emb} = -\sum\limits_{i}\sqrt{\sum\limits_{j\ne i}{\rho(r_{ij})}},
\end{equation}
where the electron density $\rho(r_{ij})$ is represented as
\begin{equation}
\rho(r_{ij}) = p_1 \exp(-p_2(r_{ij}-p_3)^2 ),
\end{equation}
with the parameters $p_1$, $p_2$, and $p_3$ to be optimized.

The magnetic energy in Eq.~\ref{eq:Etot} is given by the Stoner model for ferromagnetism~\cite{Stoner-39}
\begin{equation}
E_\mathrm{mag} =
-\frac{1}{4}\sum\limits_{i}{I_i}{m_i}^2,
\end{equation}
where $I_i$ is the Stoner exchange integral on atom $i$. The magnetic moment on atom $i$ is the difference in the number of spin-up and spin-down electrons, $m_i={N_i}^\uparrow-{N_i}^\downarrow$, after the self-consistent solution is obtained (see description of Eq.~\ref{eq:calcEbond}). 

The repulsive energy in Eq.~\ref{eq:Etot} has two main components in our model. The first is a pairwise repulsion with the same functional form as the binding energy (Eq.~\ref{eq:exponents}). In particular, we used one exponential ($k_\mathrm{max}=1$ in Eq.~\ref{eq:exponents}) for all interactions (Fe-Fe, Co-Co, and Fe-Co), which is sufficient for the pairwise repulsion as shown in previous parameterizations. The second component of the repulsive energy for our model is a short-range core repulsion of the form
\begin{equation}
E_\mathrm{core}(r_{ij}) = p_2(p_1-r_{ij})^{3}/r_{ij}\exp(p_1-r_{ij}),
\label{eq:corerep}
\end{equation}
where $r_{ij}$ is the distance between atoms $i$ and $j$. The parameters $p_1$ and $p_2$ are set to 1.5 and 100, respectively, to ensure that the atoms are strongly repulsive at a distance shorter than 1.5 \AA. 

For reproducing the elastic constant $C_{44}$ for elemental Fe, we also used an environment-dependent Yukawa repulsion for Fe-Fe interactions
\begin{equation}
E_\mathrm{Yuk} = \frac{1}{2}\sum\limits_{i,i\neq j}\frac{B}{r_{ij}}\text{exp}[-\lambda_{ij}(r_{ij}-2r_\mathrm{core})],
\end{equation}
where
\begin{equation}
\lambda_{ij} = \frac{1}{2}(\lambda_i+\lambda_j),
\end{equation}
and
\begin{equation}
\lambda_i = \lambda_0 + \Big[\sum\limits_{l\neq i}C\text{exp}(-\upsilon r_{il})\Big]^{\text{1/$m$}},
%\label{}
\end{equation}where $B$, $r_\mathrm{core}$, $\lambda_0$, $C$, $\upsilon$ and $m$ are parameters to fit; $r_{ij}$ and $r_{il}$ are the separations between atoms $i$ and $j$, or $i$ and $l$, respectively. Znam~$et$~$al.$~\cite{Znam-03} successfully used environment-dependent repulsion of this form in TiAl BOP to reproduce Cauchy pressures.

The interaction range of the BOP is limited to $r_c$ by cutoff functions 
\begin{equation}
f(r) = 
\begin{cases}
1 & \, \text{if } r < (r_{\mathrm{c}} - d_{\mathrm{c}}),\\
0 & \, \text{if }  r > r_{\mathrm{c}}, \\
\frac{1}{2} \left( \cos \biggr \langle \pi \left[ \frac{r - (r_{\mathrm{c}} - d_{\mathrm{c}}) }{d_{\mathrm{c}}} \right] \biggr \rangle + 1 \right) & \, \text{else},
\end{cases}
\label{eq:cutoff}
\end{equation}
that act on the bond integrals, embedding energy, and repulsive energy. $d_c$ is the length over which these terms go from their value at  $r=r_c-d_c$ to zero at $r=r_c$.

\section{\label{sec:Construction}Parameterization}

\subsection{\label{sec:ReferenceData}Reference data}

We fitted the Fe-Co BOP to DFT reference data. The physical model underlying the analytic BOPs allows us to work with small reference data sets. For the \mbox{Fe-Fe} and \mbox{Co-Co} interactions, we used as a reference the energy-volume \mbox{(E-V)} curves of ferromagnetic~(FM) and non-magnetic~(NM) bcc, fcc, and hcp structures which appear in the Fe-Co phase diagram~\cite{FeCoPhaDiaChap}. For the Fe-Fe interaction, we also included several sheared structures of FM bcc to get a better value of the elastic constant $C_{44}$ of elemental Fe. For the Fe-Co interaction, we used the E-V curves of FM~B2~(CsCl) as the most stable phase at 1:1 composition at low temperatures and NM B2 to cover the energy difference between FM and NM structures. We also included an FM~B32~(NaTi), which is slightly higher in energy than B2, and the bcc-based ordered L6$_0$-Fe$_3$Co~(Ti$_3$Cu) as the most stable Fe$_3$Co structure according to DFT (with marginally lower energy than D0$_3$)~\cite{DraFeCoPRL, DiazOrtiz06}. 
The DFT E-V curves that were used to parameterize the Fe-Co interaction are shown in~Fig.~\ref{fig:EVcurves}.

We performed DFT calculations using the VASP software package~\cite{vasp-1,vasp-2,vasp-3} with the projector augmented wave method (PAW) for pseudopotentials~\cite{PAW} and the PBE (Perdew-Burke-Ernzerhof)~\cite{PBE} exchange-correlation functional. We used dense Monkhorst-Pack $k$-point meshes~\cite{Kmesh} and 400 eV as energy cutoff.

\subsection{\label{sec:Parametrizationprocedure}Fitting}

We constructed the BOP for the Fe-Co in two stages. At first, using unary reference data, we parameterized the Fe-Fe and Co-Co interactions. Then, we fixed these elemental interactions and parameterized the Fe-Co interaction with binary reference data. We fit the parameters of the Fe-Fe, Co-Co, and Fe-Co interactions to the corresponding E-V curves using the least square method with higher weights for the most stable phases of Fe, Co, and Fe-Co. We optimized the parameters of our model with the \mbox{BOPcat} software package~\cite{bopcat} using the Levenberg–Marquardt method~\cite{LeastSq-1,LeastSq-2}.

We chose initial guesses for the bond integrals (Eq.~\ref{eq:exponents}) for \mbox{Fe-Fe} and \mbox{Co-Co} interactions based on a previous parameterization~\cite{CoBOPApa}. For \mbox{Fe-Co}, we took the initial guess as the average of the optimized Fe-Fe and Co-Co bond integrals because the TB bond integrals of the \mbox{Fe-Co} dimer are between those of the \mbox{Fe-Fe} and \mbox{Co-Co} dimers (see discussion in Ref.~\cite{JenkeDimers}). 

The number of valence electrons and the Stoner exchange integral are continuous parameters in the TB/BOP~\cite{DraPettMagBOP11} formalism. We chose the number of valence $d$ electrons as 6.98 for Fe and 8.20 for Co to reproduce the FM bcc and FM hcp as the ground state structures, respectively. By setting the Stoner exchange integrals to 0.751 eV and 0.841 eV for Fe and Co, we captured energy differences between magnetic and non-magnetic structures for elemental Fe and Co (see Fig.~2 in Supplemental Material~\cite{MySuppl}).

We optimized the initial parameters of energy contributions in Eq.~\ref{eq:Etot} to achieve the best agreement with the reference data. The resulting parameters are compiled in Table 1 in the Supplemental Material~\cite{MySuppl}. Figure~\ref{fig:bondint} shows the optimized bond integrals $dd\sigma$, $dd\pi$, and $dd\delta$ as a function of distance $R$ (see Eq.~\ref{eq:exponents}) for the \mbox{Fe-Co} interaction compared with the bond integrals obtained by projecting the DFT eigenstates of the Fe-Co dimer onto an orthogonal TB basis~\cite{JenkeDimers}. The Fe-Co bond integrals optimized for bulk reference data are close to the bond integrals for the dimer. They differ visually because the surrounding atoms screen the bond between two atoms in a solid~\cite{ScreBondIn}, which is not the case for a dimer. 
The resulting bond integrals for the \mbox{Fe-Fe} and \mbox{Co-Co} interactions are shown in Fig.~1 in the Supplemental Material~\cite{MySuppl}.

\begin{figure}[t]
\begin{center}
\includegraphics[width=0.445\textwidth]{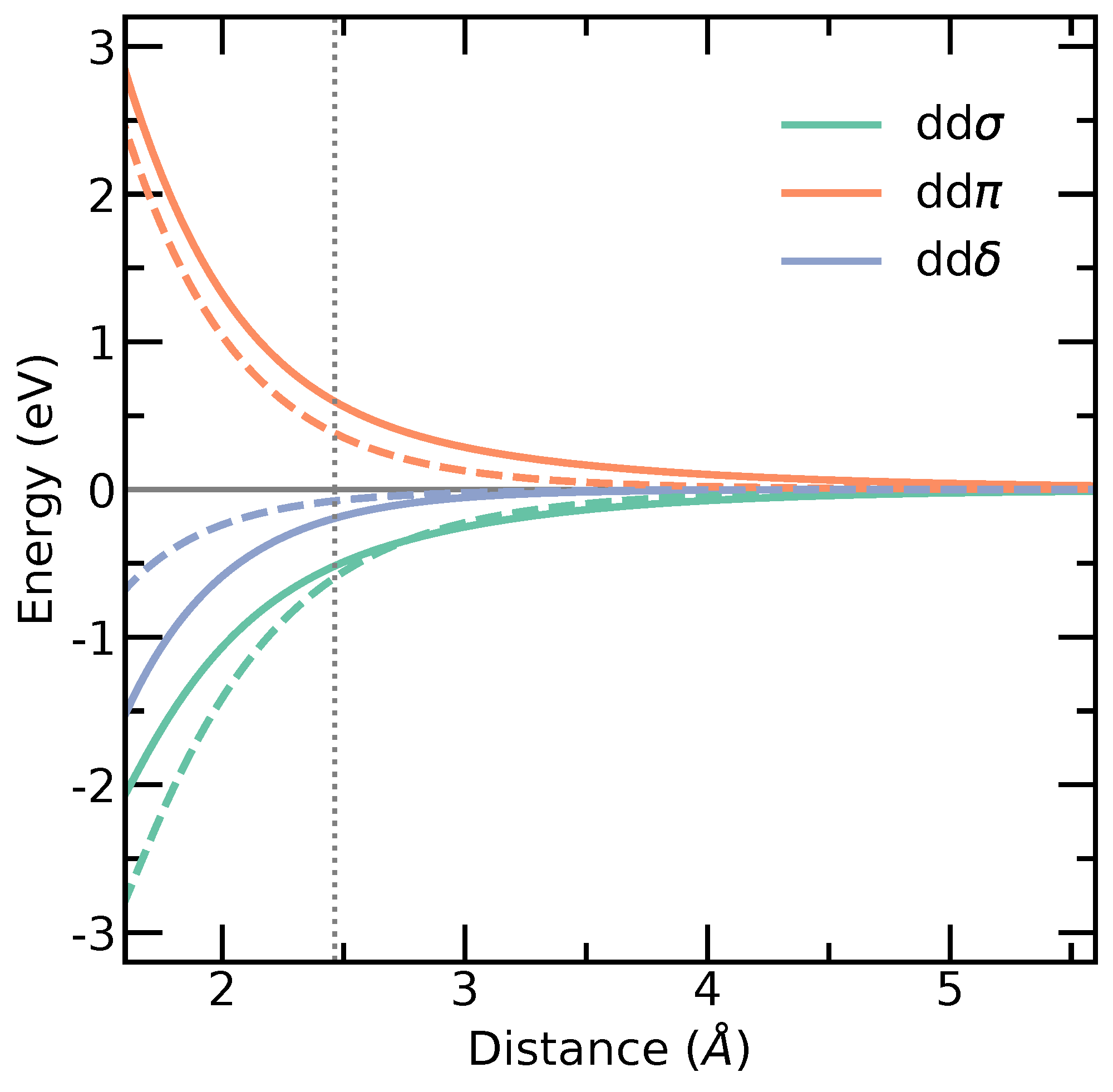}
%\hspace{-0.5 cm}
%\vspace{7cm} %\includegraphics[width=0.9\columnwidth]{}
\caption{Bond integrals for \mbox{Fe-Co} as obtained from downfolding the DFT wavefunction of a dimer onto an orthogonal TB basis~\cite{JenkeDimers} (dashed lines) and after our BOP optimization (solid lines) as a function of distance. A dotted grey line indicates the first nearest neighbor distance in ferromagnetic (FM) \mbox{B2 FeCo}.}
\label{fig:bondint}
\end{center}
\end{figure}
\begin{figure}[t]
\begin{center}
\includegraphics[width=0.45\textwidth]{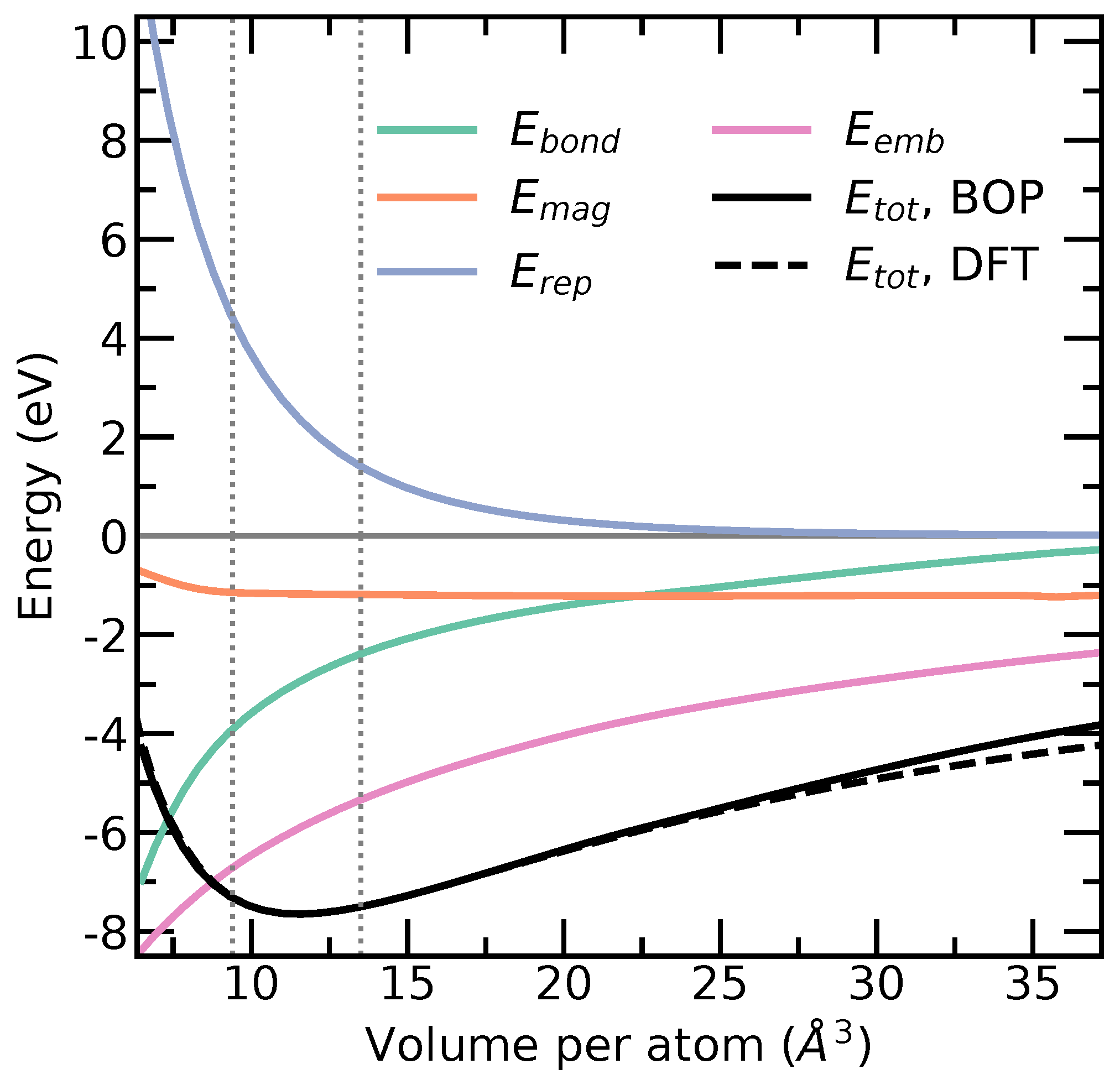}
\caption{Contributions to the total binding energy (Eq.~\ref{eq:Etot}) of the parameterized BOP for ferromagnetic (FM) B2 FeCo.  The BOP total binding energy curve is compared with corresponding DFT results. The dotted grey lines indicate the range of volumes that we considered in the parameterization.}
\label{fig:allcontrib}
\end{center}
\end{figure}

Figure~\ref{fig:allcontrib} shows the contributions of the different terms to the total binding energy (Eq.~\ref{eq:Etot}) for FM B2 FeCo for the optimized BOP. The embedding energy $E_\mathrm{emb}$ that mimics the binding from $s$ electrons and the bond energy $E_\mathrm{bond}$ from the $d$ electrons are attractive for all volumes and counteracted by the repulsive energy $E_\mathrm{rep}$. The magnetic energy $E_\mathrm{mag}$ delivers a nearly constant energy gain that vanishes at small volumes when the magnetic moment starts to collapse. The core repulsion (Eq.~\ref{eq:corerep}) is short-range and not seen in this graph. The total binding energy agrees with DFT results for volumes far beyond those we used as a reference. This is due to the underlying physical model, which leads to high transferability.

In the following section, we validate the Fe-Co BOP by computing the electronic structure, structural stability, structural transformations, elastic constants, phonons, point and planar defects, and formation energies of Fe-Co alloys. The elemental Fe and Co BOP validation are compiled in the Supplemental Material~\cite{MySuppl} (see, also, Refs.~\cite{fromsuppl-1-1-lizarraga2017cobphon, fromsuppl-1-2-cohen06dftPhonBccFe, fromsuppl-1-3-VacFormAstaDFT15} therein). It is worth noting that our Fe BOP shows improvements compared to existing BOP parameterizations~\cite{MroFePRL11,Ford-14}, e.g., for the fcc and hcp structures.

\section{\label{sec:Performance}Validation}

\subsection{\label{sec:ElecStruc}Electronic structure and magnetism of B2 FeCo}

\begin{figure}[t]
\centering\includegraphics[width=0.48\textwidth]{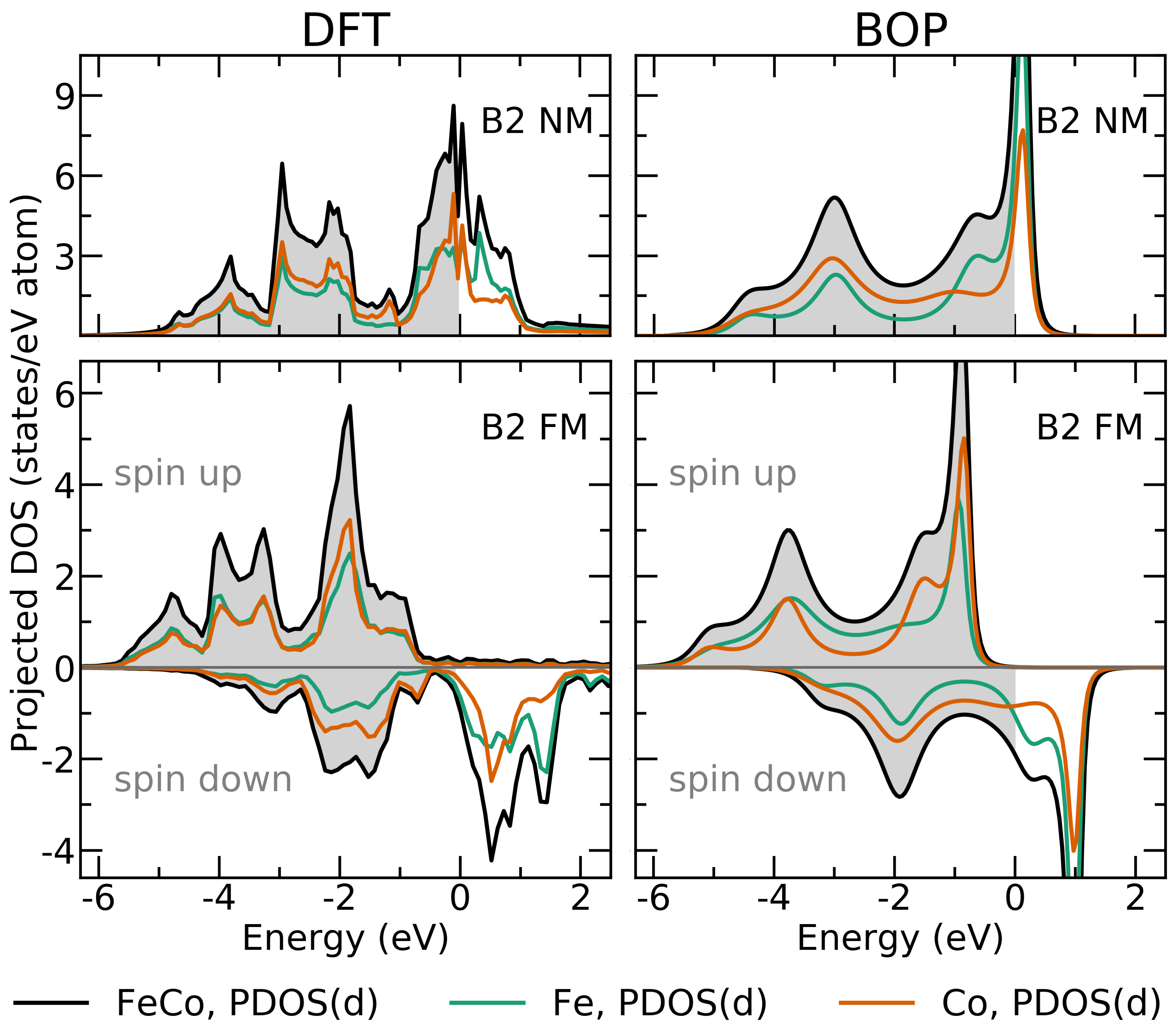}
\caption{Electronic DOS of non-magnetic (NM) and ferromagnetic (FM) B2 FeCo obtained with DFT (left) and BOP (right). For the DFT-DOS, the projection on $d$ orbitals is plotted for consistent comparison with $d$-valent BOP. The Fermi energy is taken as 0~eV.}
\label{fig:electronDOS}
\end{figure}

The BOP model is based on a coarse-grained description of the electronic structure. Therefore a direct and intuitive test compares the electronic DOS predicted with BOP and the DOS obtained with DFT. In Fig.~\ref{fig:electronDOS}, we compare them for B2 FeCo, the ground state structure at 1:1 composition, for NM and FM configurations. The DOS is split into the contributions from Fe and Co atoms in both BOP and DFT. For a consistent comparison with our $d$-valent BOP, we consider only the projection of the DFT-DOS on $d$ orbitals.

The BOP-DOS is well-matched with the DFT-DOS for both NM and FM configurations. It is smoother overall due to the chosen low number of moments (see Sec.~\ref{sec:abop}), but it captures the positions and height of the main features and the bandwidth. The BOP captures the peak at the Fermi level for the NM configuration, fulfilling the Stoner criterion and stabilizing the FM B2 phase. It also correctly describes the relative weight of the Fe and Co atoms. In general, an accurate description of the DOS means that the BOP captures the physics of magnetism in the interaction of Fe and Co atoms. We can obtain the BOP-DOS that agrees better with DFT if we use not nine but more moments. However, this will increase the computational cost of calculations. Fig.~4 in the Supplemental Material~\cite{MySuppl} shows an example of the BOP-DOS for FM B2 FeCo computed with 25 moments that reproduces more features of the DFT-DOS.

The electronic DOS of magnetic B2 FeCo obtained with our BOP (Fig.~\ref{fig:electronDOS}) results in magnetic moments of 3.01$\mu_B$ for Fe and 1.81$\mu_B$ for Co atoms. For elemental bcc Fe and bcc Co, our BOP gives 2.88$\mu_B$ and 1.80$\mu_B$, respectively. Thus, the magnetic moments are almost the same for Co atoms, while they are significantly higher for Fe atoms in the B2 phase compared to elemental bcc. This behavior of the Fe and Co magnetic moments is consistent with DFT calculations. The DFT results in B2 are 2.80$\mu_B$ for Fe atoms and 1.81$\mu_B$ for Co. For elemental bcc, DFT magnetic moments are 2.24$\mu_B$ and 1.82$\mu_B$, respectively~\cite{DFTmagmomsB2andpurebccFeandCo}.

\subsection{\label{sec:Stability}Energy-volume curves of ordered Fe-Co alloys}

\begin{figure}[t]
\centering\includegraphics[width=0.47\textwidth]{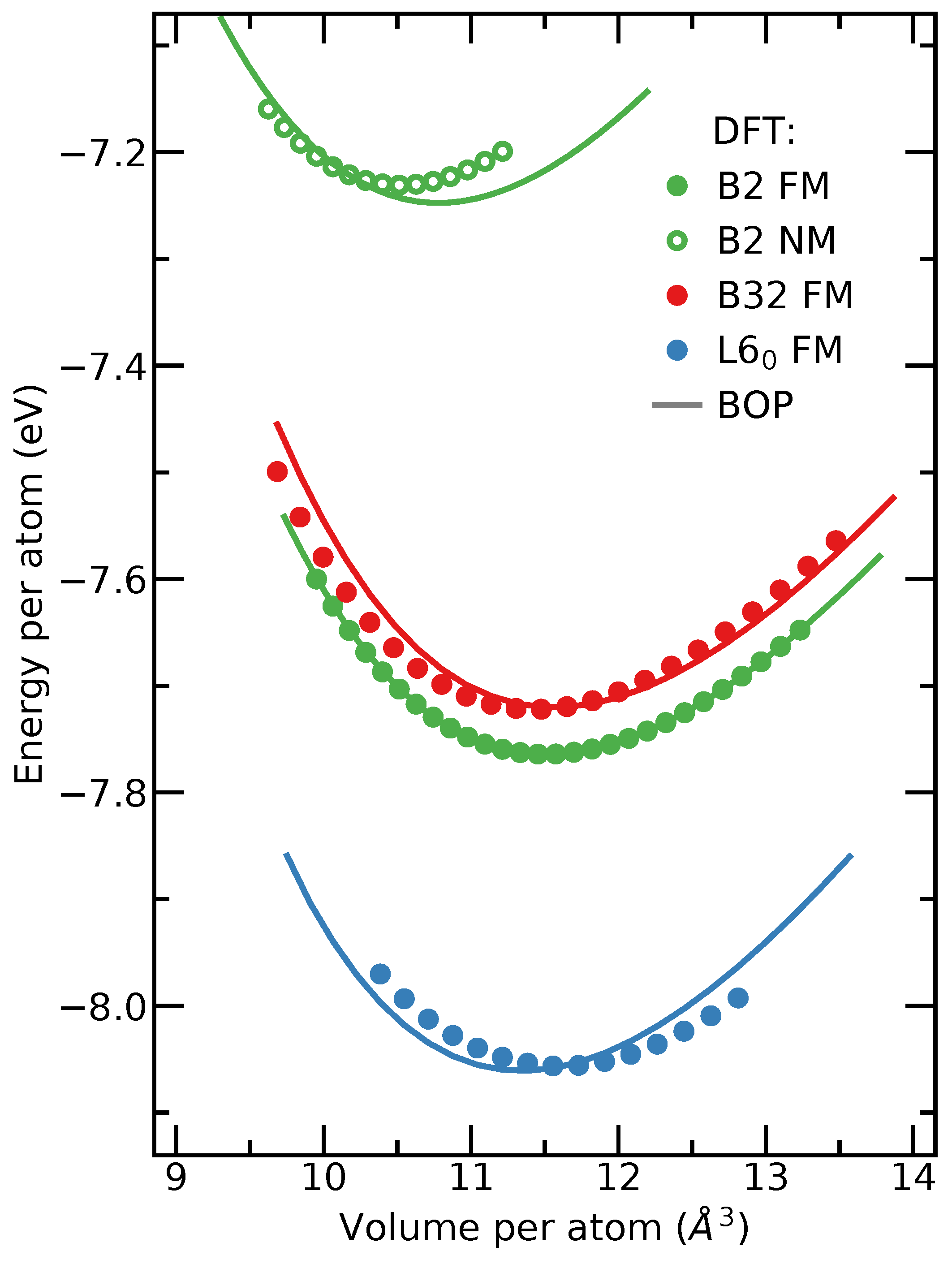}
\caption{Comparison of energy-volume (E-V) curves of Fe-Co BOP (lines) and all DFT reference data points we used in the parameterization of Fe-Co interaction (circles).}
\label{fig:EVcurves}
\end{figure}

We further validate the Fe-Co BOP by assessing the total binding energy $E_\mathrm{tot}$ (Eq.~\ref{eq:Etot}). Figure~\ref{fig:EVcurves} compares BOP and DFT E-V curves for the complete set of Fe-Co reference data we used.
The BOP perfectly reproduces the DFT E-V curve for FM B2, the most stable phase at 1:1 composition. While the E-V curve of NM B2, which has higher energy, also agrees well, given that we gave it lower weights during fitting. The basis for capturing this energy difference between magnetic and non-magnetic B2 is the accurate description of the underlying electronic structure, as shown in Fig.~\ref{fig:electronDOS}.

\subsection{\label{sec:Alloys}Structural stability across chemical compositions}

To assess the reliability of the Fe-Co BOP for various chemical compositions, we determined the convex hull of the formation energies at 0~K. The convex hull of the Fe-Co system, in addition to B2, exhibits a dense set of ordered bcc-based stable structures for Fe-rich compositions in regular composition steps of 1/16~\cite{DraFeCoPRL}. In contrast, all structures on the Co-rich side lie above or close to the tie line connecting B2 and hcp Co.

To compare our BOP with DFT, we computed energies of formation of the Fe-Co binary alloys as
\begin{equation}\label{eq:dH-FeCo}
E_f = \frac{E(\mathrm{Fe}_x\mathrm{Co}_y)  
             - x E(\mathrm{Fe}) - y E(\mathrm{Co})}{x+y},
\end{equation} where $E(\mathrm{Fe}_x\mathrm{Co}_y)$ is the total energy of a given alloy's supercell and $E(\mathrm{Fe})$, and $E(\mathrm{Co})$ are total energies per atom of the elemental ground states bcc Fe and hcp Co. The crystal structures from the convex hull are given in the Appendix of Ref.~\cite{DiazOrtiz06}.

Figure~\ref{fig:convexhull} compares the energies of formation calculated with DFT~\cite{DiazOrtiz06}, BOP, and MEAM~\cite{Choi2017meam}.
\begin{figure}[t]
\centering\includegraphics[width=0.482\textwidth]{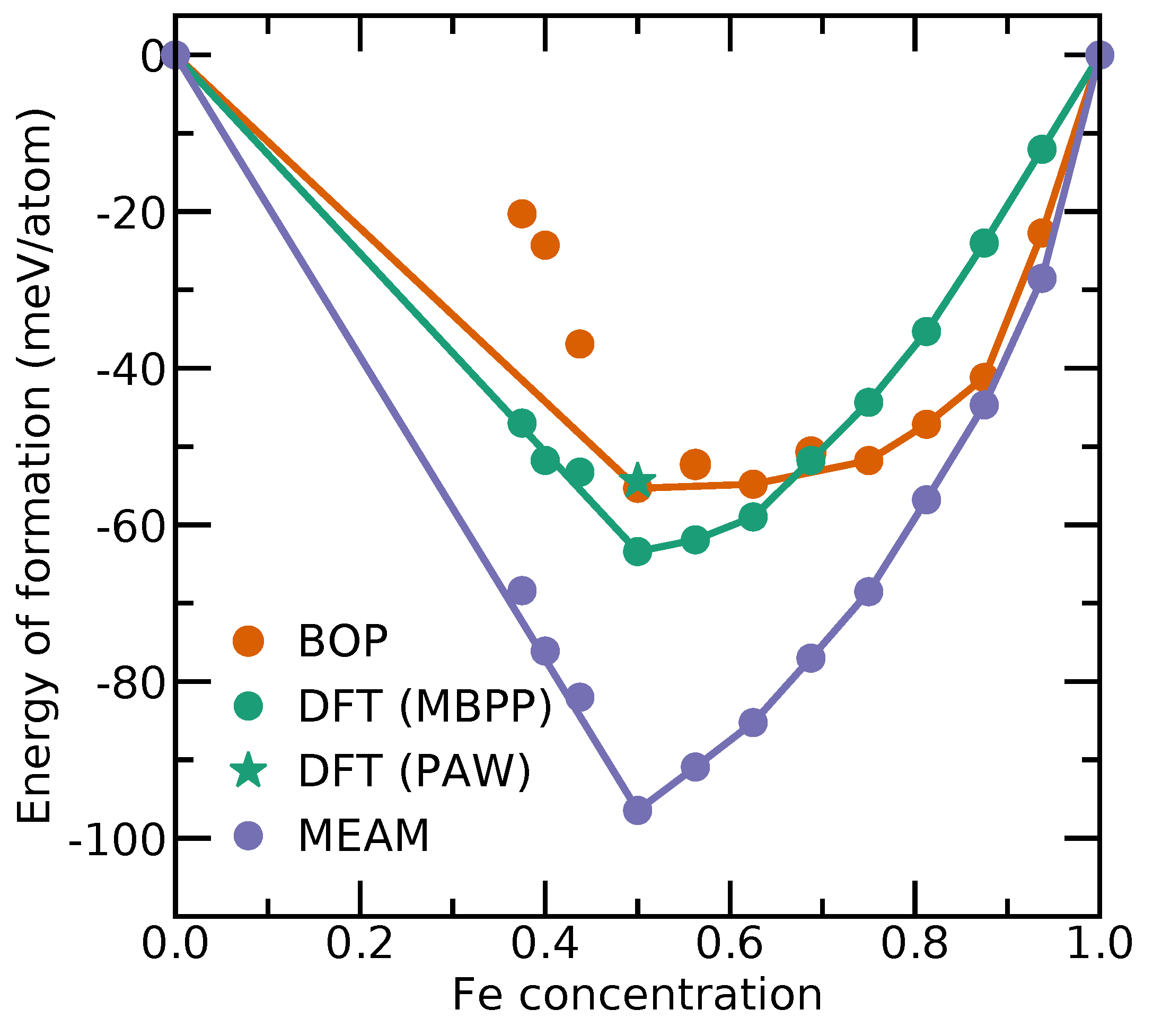}
\caption{Energies of formation for ordered bcc-based Fe-Co structures (circles) at various concentrations calculated with BOP, DFT~\cite{DiazOrtiz06}, and MEAM~\cite{Choi2017meam}. Lines connect the stable structures forming a convex hull.}
\label{fig:convexhull}
\end{figure} 
The BOP reproduces the dense set of stable structures for Fe-rich compositions with an accuracy of approximately 10~meV/atom. The~Fe$_9$Co$_7$ and Fe$_{11}$Co$_5$ structures are not on the BOP convex hull but close within less than 3~meV, which seems remarkable as only two stable structures, B2 and L6$_0$, were included in the reference data. DFT and BOP B2 energies differ because the DFT data was obtained with the mixed-basis pseudopotential (MBPP) method, whereas we fitted the BOP to DFT obtained with PAW (cf. Table~\RNum{2} in Ref.~\cite{DiazOrtiz06}). 
The difference between MEAM and DFT is significant and may be due to the choice of reference data in the construction of the MEAM.

\subsection{\label{sec:phasestab}Structural stability of disordered Fe-Co alloys}

Fe-Co at 1:1 composition transforms from ordered B2 to disordered bcc (A2) at around 1000~K and further to disordered fcc (A1) above 1200~K~\cite{FeCoPhaDiaChap}. To assess whether our BOP may reproduce these phase transitions, we determined the relative stability of A1, A2, and B2 at 1:1 composition at 0~K as a precursor for predicting the phase transitions~\cite{TiTaChakrob}. We considered FM and NM configurations as the A2-A1 transition temperature coincides with the Curie temperature. We represented the bcc and fcc solid solutions by special quasi-random structures (SQS)~\cite{SQS} with 16-atom cells from Ref.~\cite{sqsbcc} and Ref.~\cite{sqsfcc}, respectively.

\begin{figure}[t]
\begin{center}
\includegraphics[width=0.485\textwidth]{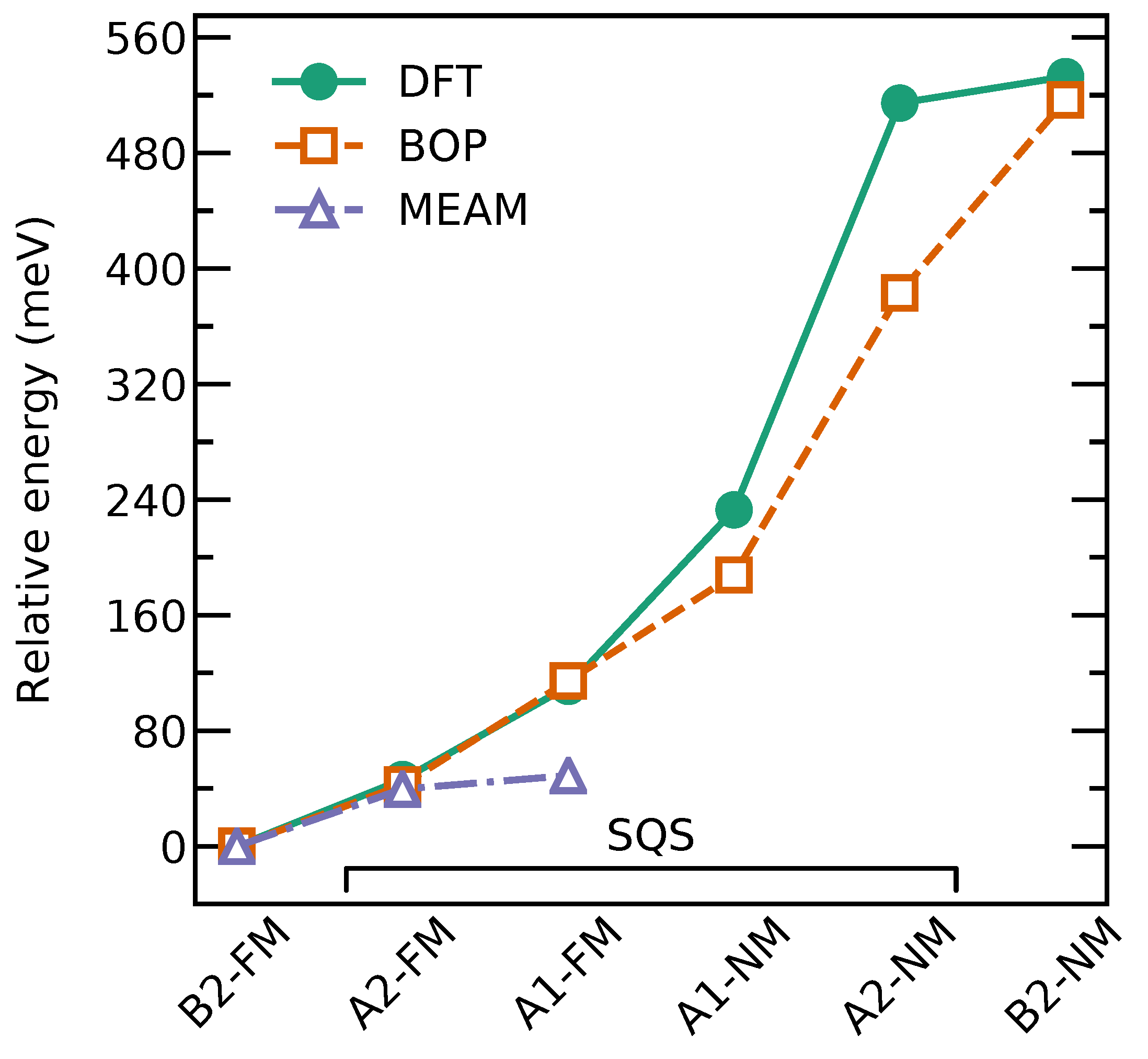}
\caption{
%\textcolor{blue}
{Structural stability of B2, disordered bcc (A2), and fcc (A1) phases in ferromagnetic (FM) and non-magnetic (NM) configurations relative to FM B2 computed with BOP, DFT, and MEAM.}
}
\label{fig:phasestab}
\end{center}
\end{figure}

The comparison to DFT (and also MEAM) results in Fig.~\ref{fig:phasestab} shows that the BOP accurately predicts the sequence of structural stability from ordered B2-FM to disordered A2-FM to disordered A1-FM. It also predicts the correct sequence for the energetically higher NM configurations with the most significant deviation for A2-NM. Thus, we see that BOP has strong predictive power for disordered phases that were not part of the reference data. Our results also confirm the conclusions of Abrikosov~$et$~$al.$~\cite{AbrikFeCoPRB} made with DFT using coherent potential approximation (CPA) that magnetism is crucial for stabilizing B2 relative to disordered phases. We should note that the MEAM results in Fig.~\ref{fig:phasestab} are given only for FM configuration as it was fitted to FM reference data, and magnetism is not included in this potential explicitly.

\subsection{\label{sec:Elastic}Elastic constants of B2 FeCo}

To test the mechanical properties, we compared the bulk modulus ($B$), elastic constants ($C_{11}$, $C_{12}$, $C_{44}$), and tetragonal shear constant ($C'$) computed with BOP to values calculated with DFT, embedded atom method (EAM)~\cite{EAM} and modified EAM (MEAM)~\cite{MEAM-1,MEAM-2,MEAM-3} potentials as well as available experimental data (see Tab.~\ref{tab:elconst}).
\begin{table}[b]
\caption{\label{tab:elconst}
Bulk modulus and elastic constants for ferromagnetic (FM) B2 FeCo obtained using BOP, DFT, experiment, and empirical interatomic potentials. All values are in GPa.
}
\begin{ruledtabular}
\begin{tabular}{cccccc}
 &$B$&$C_{11}$ &$C_{12}$ &$C_{44}$& $C'$\\
\hline
BOP\footnotemark[1] & 190 & 268 & 153 & 85 & 58\\
DFT\footnotemark[1] & 191 & 271 & 155 & 128 & 58\\
DFT\footnotemark[2] & 188 & 263 & 152 & 108 & 56\\
DFT\footnotemark[3] & 189 & 259 & 154 & 131 & 53\\
MEAM\footnotemark[4] & 196 & 268 & 179 & 138 & 27\\
EAM\footnotemark[5] & 187 & 226 & 168 & 136 & 29\\
MEAM\footnotemark[6] & 188 & 261 & 151 & 127 & 55\\
exp. & 188\footnotemark[7] &  &  &  & 61\footnotemark[8]\\
\end{tabular}
\end{ruledtabular}
\footnotetext[1]{This work}
\footnotetext[2]{Ref. \cite{rahaman2013phd}}
\footnotetext[3]{Ref. \cite{Jain2013materialsproject}}
\footnotetext[4]{Ref. \cite{Choi2017meam}}
\footnotetext[5]{Ref. \cite{Li2021eam}}
\footnotetext[6]{Ref. \cite{Muralles2022meam}}
\footnotetext[7]{Ref. \cite{belousov2009expB}, (measured at 300 K)}
\footnotetext[8]{Ref. \cite{clark2008ExpCprime}, (measured at 61 K)}
\end{table}
The BOP values for bulk modulus, $C_{11}$, $C_{12}$, and $C'$ elastic constants agree with the DFT results. BOP underestimates $C_{44}$ due to the absence of sheared Fe-Co structures in the reference data. 

\subsection{\label{sec:Phonons}Phonon spectrum of B2 FeCo}

\begin{figure}[t]
\centering\includegraphics[width=0.456\textwidth]{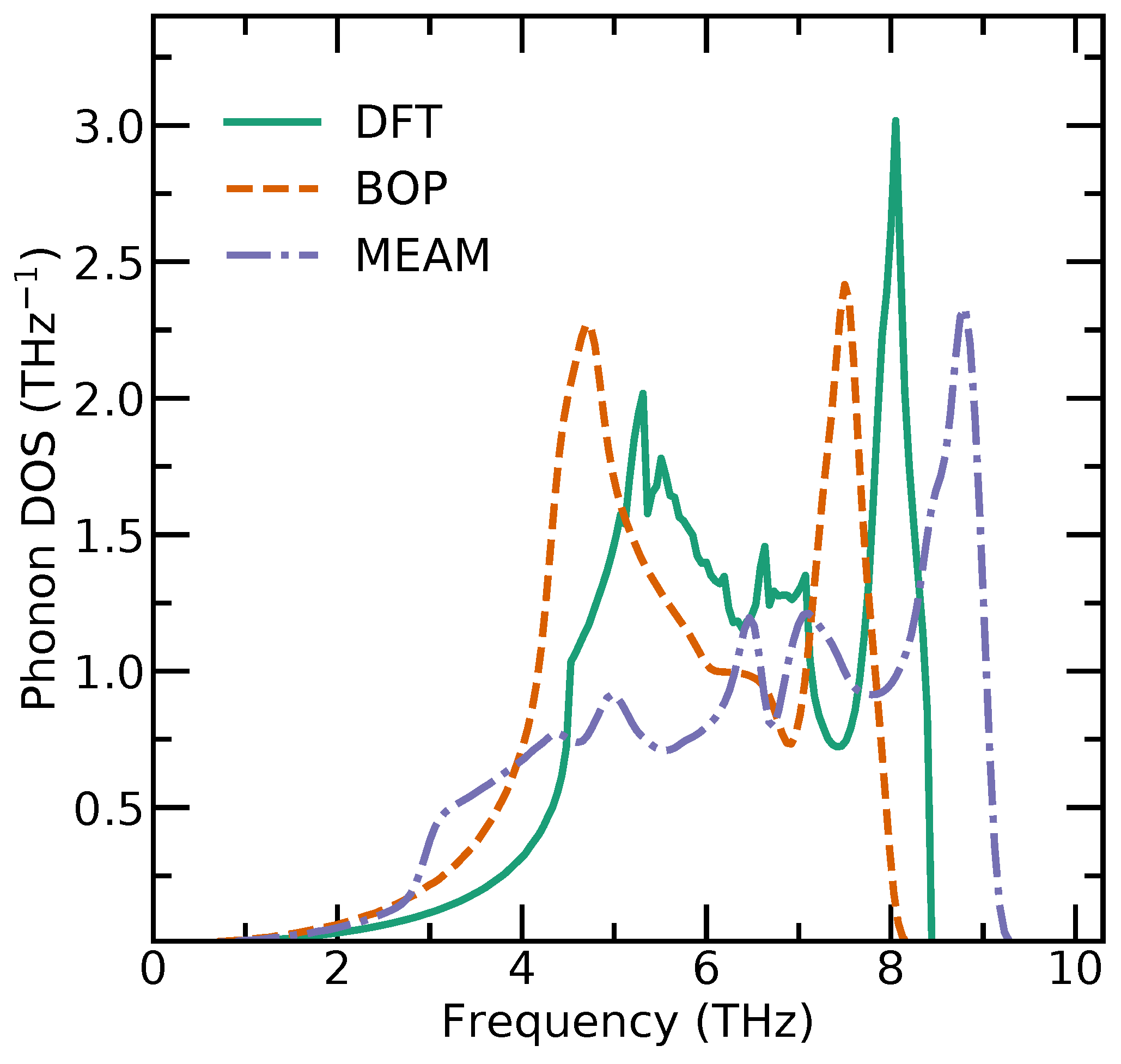}
\caption{Phonon DOS of ferromagnetic (FM) B2 FeCo obtained using DFT, BOP and MEAM~\cite{Choi2017meam}.}
\label{fig:phononDOS}
\end{figure}

Existing TB/BOP models for Fe~\cite{MroFePRL11,MadsenFe11,Ford-14} have captured magnetic~\cite{Wang-2019} and structural~\cite{Wang-2022} phase transitions at finite temperatures. Interatomic potential should accurately describe lattice vibrations to achieve transferability beyond the 0~K reference data. In Fig.~\ref{fig:phononDOS}, we compare the phonon DOS of FM B2 FeCo calculated with BOP, DFT, and MEAM~\cite{Choi2017meam}. We used the phonopy package to compute the phonon DOS~\cite{phonopy}.
BOP reproduces the main features of the DOS with two prominent peaks and an intermediate peak in between, while the MEAM produces a different shape. Our BOP slightly underestimates the frequency width of the DOS, but overall the BOP predictions are convincing, given that the reference data did not include forces.

\subsection{\label{sec:PointDefects}Point defects in B2 FeCo}

Point defects play a crucial role in Fe-Co and largely determine the stability of the ordered B2 phase~\cite{FahnleAntisite}. For this reason, we calculated the formation energies of vacancies and antisite defects for B2 with a 1:1 composition at 0~K with our BOP. We computed approximate formation energies following derivations of Meyer and Fähnle for antistructure-type systems (Eq.~A7 in Ref.~\cite{FahnleDefFormEnFormula1999}). They assumed that for such systems, the formation energies of antisites of different species are equal. In the Supplemental Material~\cite{MySuppl} (see, also, Refs.~\cite{fromsuppl-2-1-ChemPotBOP, fromsuppl-2-2-B2DefForEnEq} therein), we calculated the formation energies differently and confirmed that our BOP predicts the assumed antistructure-type of B2 FeCo.

\begin{figure}[t]
%\captionsetup{width=1.0\textwidth}
\centering\includegraphics[width=0.455\textwidth]{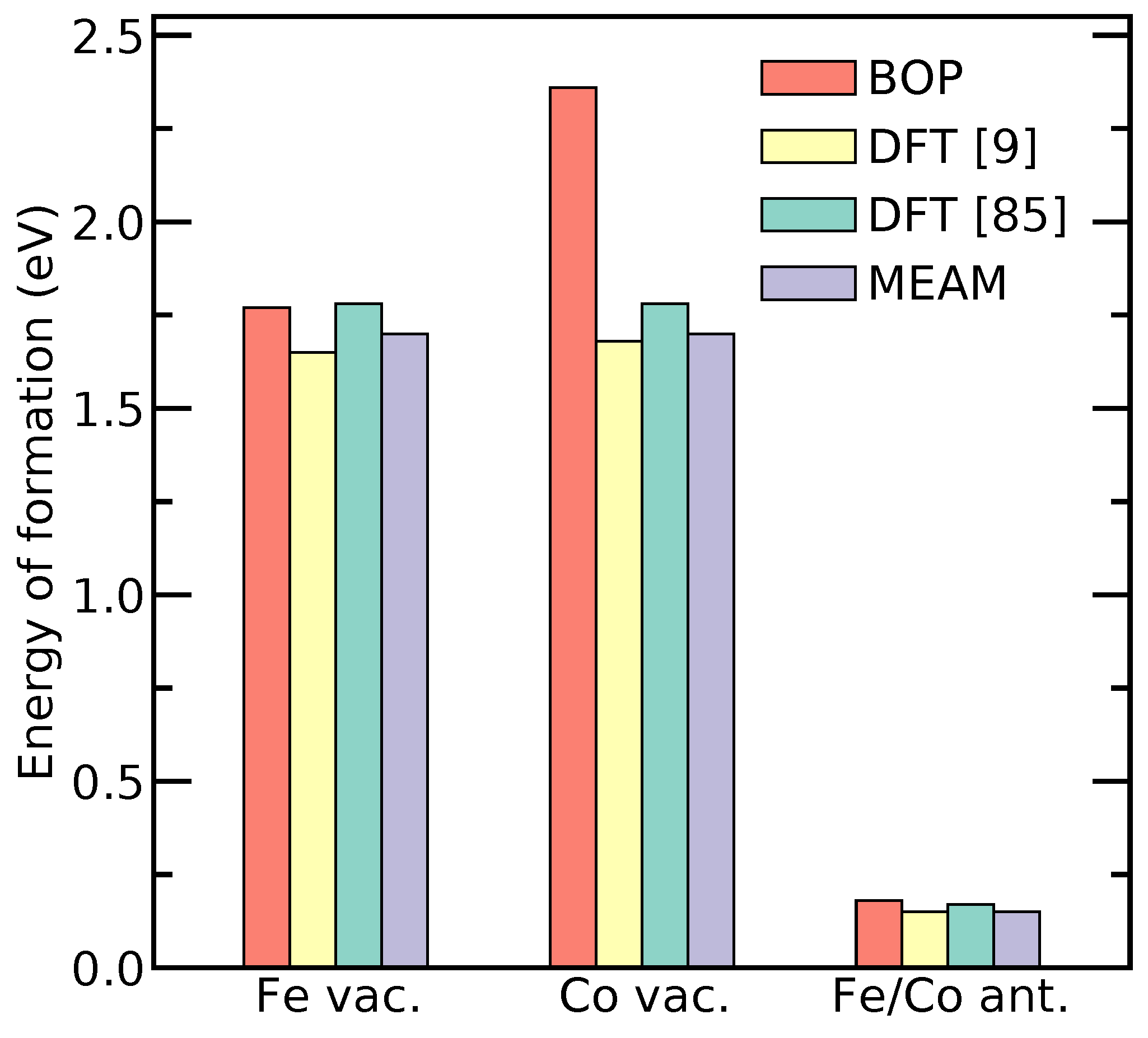}
\caption{Defect formation energies for Fe and Co vacancies and antisites in ferromagnetic (FM) B2 FeCo at 1:1 composition at 0~K computed using BOP, DFT~\cite{FahnleAntisite,KrchmarPRB06} and MEAM~\cite{Choi2017meam}. The formation energies of Fe and Co antisites are the same (according to Eq.~A7 in Ref.~\cite{FahnleDefFormEnFormula1999}; see details in section~\ref{sec:PointDefects}).}
\label{fig:defects}
\end{figure}
Figure~\ref{fig:defects} compares the formation energies of vacancies and antisite defects obtained using BOP, DFT~\cite{FahnleAntisite,KrchmarPRB06}, and MEAM~\cite{Choi2017meam}. Even though the BOP value for a Co vacancy is slightly higher than the DFT, overall, BOP correctly predicts that the formation energies of vacancies are much higher than of antisite defects, making the latter the dominant defects in B2 FeCo. The small but still positive formation energy of antisite defects prevents the spontaneous exchange of Fe and Co atoms and hence stabilizes the B2 phase relative to disordered bcc. The formation energies of defects were not part of the reference data for our BOP, and their correct reproduction demonstrates the transferability of the potential. 

\subsection{\label{sec:Transpath}{B2-L1$_{0}$ transformation path}}

DFT predicts that the tetragonal distortion by increasing the $c/a$ ratio of the lattice parameters of B2~FeCo will produce magnetic properties desirable for storage devices: high magnetic saturation moment and giant magnetic anisotropy~\cite{giantmaganis}. Such a distortion is feasible in practice, for example, in L1$_{0}$~FeCo, which can be grown as a thin film on Cu substrates~\cite{prlbainpathandgrowth}. For this reason and to validate our BOP against atomic environments that differ from bcc, we computed the transformation path from bcc-based B2 to fcc-based L1$_{0}$ by continuous variation of the $c/a$ ratio.

\begin{figure}[t]
\begin{center}
\includegraphics[width=0.456\textwidth]{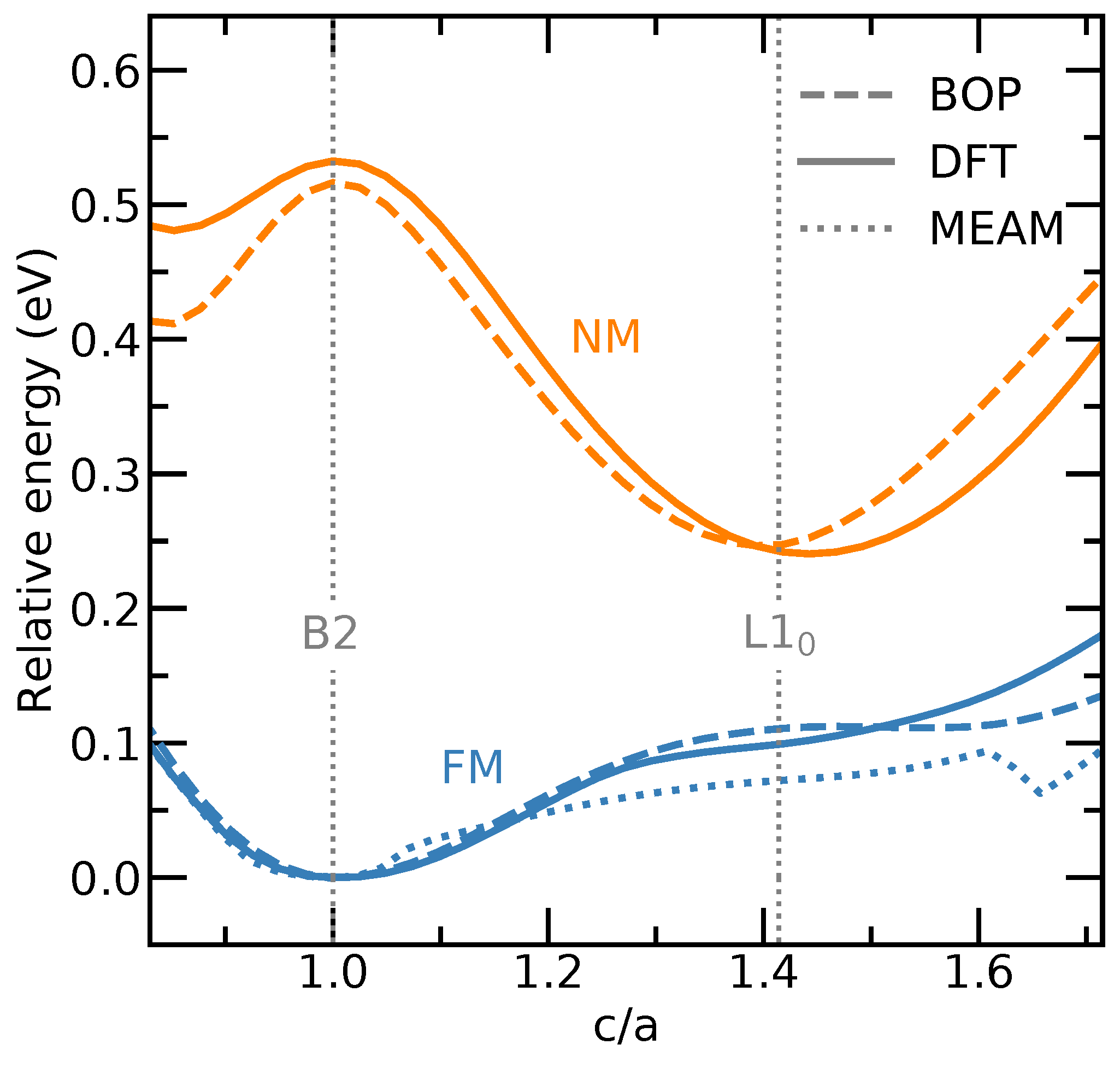}
\caption{
%\textcolor{blue}
{Transformation path from bcc-based B2 to fcc-based L1$_{0}$ phase for ferromagnetic (FM) and non-magnetic (NM) configurations obtained with BOP, DFT, and MEAM~\cite{Choi2017meam}.}
}
\label{fig:transpath}
\end{center}
\end{figure}
The results of BOP, DFT, and MEAM~\cite{Choi2017meam} for FM and NM (not applicable for MEAM; see~\ref{sec:phasestab}) configurations are compiled in Fig.~\ref{fig:transpath} using the standard Bain path notation with $c/a=1$ for B2 and $c/a=1.41$ for L1$_{0}$. Overall the BOP precisely reproduces the DFT transformation path for both FM and NM configurations. The agreement for the NM case is particularly striking because even the only reference structure along the transition path, B2, had a low weight during the fitting. This finding highlights the predictive power of the Fe-Co BOP to atomic environments beyond the reference data that included only structures with $c/a = 1$.

\subsection{\label{sec:GBs}Grain-boundary energies}

To test atomic environments with bond angles and nearest-neighbor coordination different from ideal crystal structures, we investigated three typical grain boundaries (GBs) in B2 FeCo. We considered the $\Sigma$5[001](001) and $\Sigma$3[011](011) twist GBs as well as the $\Sigma$3[011](1-11) tilt GB with rotation angles of 36.87°, 70.53°, and 70.53°, respectively. We constructed the corresponding supercells with 120, 144, and 72 atoms and six B2 unit cell distances between two adjacent GB planes with the aimsgb package~\cite{gbpackage}.
We computed the formation energy of the GBs, $\gamma_\mathrm{GB}$, by DFT, BOP, and MEAM with complete relaxation  of atomic positions as
\begin{equation}\label{eq:EGB}
\gamma_\mathrm{GB} = \frac{E_\mathrm{GB} - E_\mathrm{bulk}}{2A_\mathrm{GB}},
\end{equation}
where $E_\mathrm{GB}$ is the energy of a supercell with GB of area $A_\mathrm{GB}$ and $E_\mathrm{bulk}$ is the energy of a bulk supercell with the same number of atoms. 

\begin{figure}[t]
\centering\includegraphics[width=0.46\textwidth]{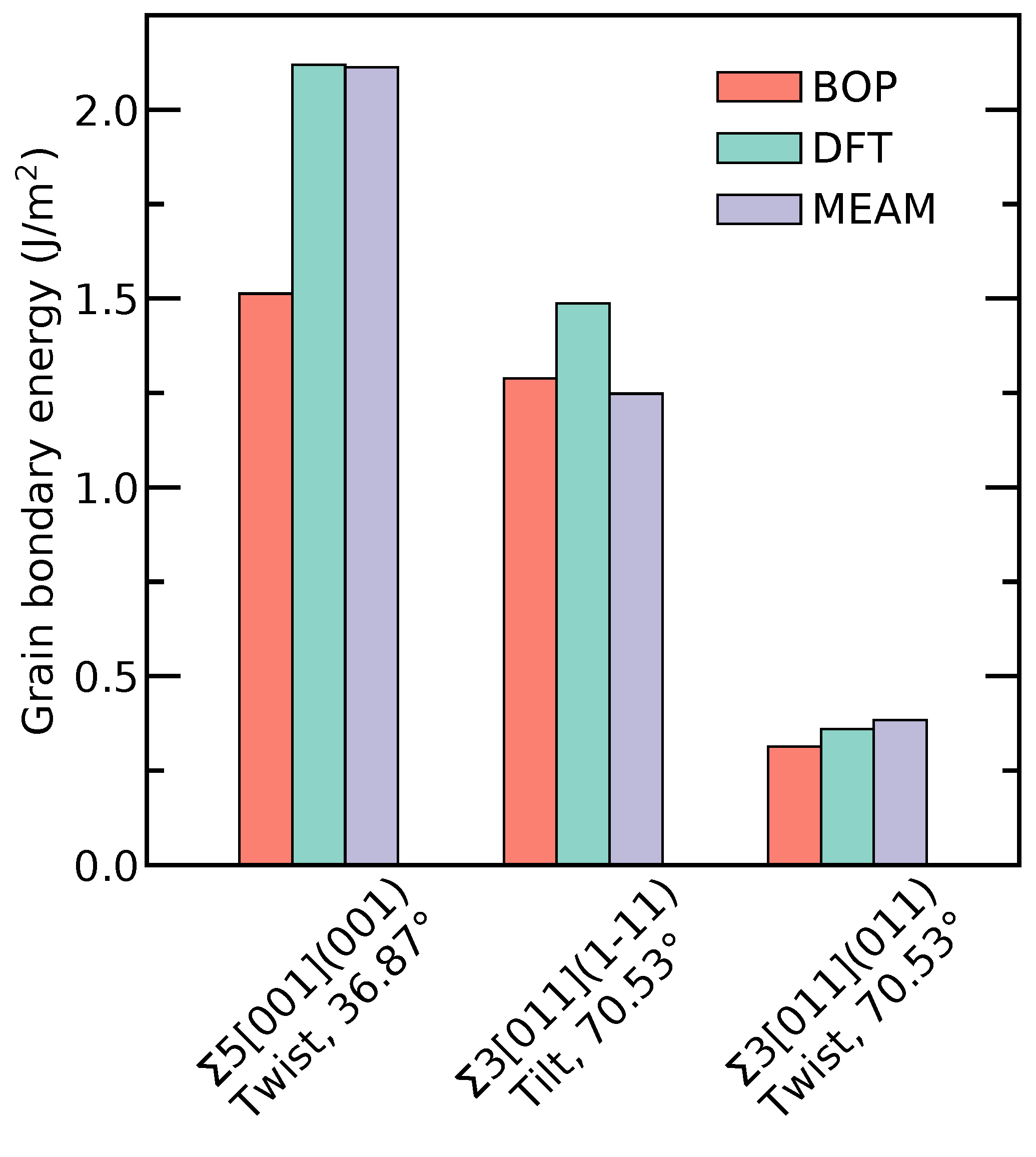}
\caption{Energies of two twist and one tilt grain boundaries (GBs) in the ferromagnetic (FM) B2 phase computed using BOP, DFT, and MEAM~\cite{Choi2017meam}.}
\label{fig:gb}
\end{figure} 

The comparison in Fig.~\ref{fig:gb} shows that MEAM~\cite{Choi2017meam} and BOP reproduce the energetic ordering of the GBs. BOP accurately captures the energies of the $\Sigma$3-twist and the $\Sigma$3-tilt GBs while slightly underestimating the $\Sigma$5-twist GB. As we see, our BOP shows considerable transferability to complex atomic environments, such as GBs.

\section{Conclusions}

We developed an analytic bond-order potential for Fe-Co alloys that includes an explicit treatment of magnetism. We use a $d$-valent orthogonal tight-binding Hamiltonian in two-center approximation and employ an embedding function to account for the $s$ electrons. The functional form is physically transparent and, for this reason, requires only a small set of reference data. At the same time, the underlying physics ensures robust predictions of Fe-Co properties that we did not include in the fit. We demonstrate the transferability of our potential to various material properties: structural stability of ordered and disordered phases, elastic constants and phonons, point defects, structural transformations, and planar defects. Due to the explicit treatment of magnetism, our BOP reproduces the main features of the DFT electronic structure for magnetic and non-magnetic phases. Further, the BOP reproduces the dense sequence of stable phases for Fe-rich Fe-Co alloys with an accuracy of about 10~meV to DFT results and the B2 stability against disordered phases provided by magnetism. The analytic BOP for Fe-Co paves the way to atomistic simulations of Fe-Co alloys with a reliable treatment of magnetism at length and time scales inaccessible with DFT.

\section*{Acknowledgment}
We acknowledge financial support from the International Max-Planck Research School SurMat and the Wilhelm and Günter Esser Foundation. Part of this work was supported by the German Research Foundation (DFG) through project C1 of the collaborative research center SFB/TR 103 (190389738) as well as by the DFG and the French National Research Agency within the DFG-ANR project MAGIKID (316673557). 
We acknowledge Yury Lysogorskiy for technical support and Isabel Pietka for carefully reading the manuscript.

\ifarXiv
    

\section*{Supplementary material}

# Supplemental Material: "Magnetic bond-order potential for iron-cobalt alloys"

Aleksei Egorov[1], Aparna P.A. Subramanyam[1], Ziyi Yuan[2], Ralf Drautz[1], and Thomas Hammerschmidt[1]

[1]ICAMS, Ruhr-Universität Bochum, Universitätstr. 150, 44780 Bochum, Germany
[2]Department of Materials, University of Oxford, Parks Road, Oxford OX1 3PH, United Kingdom

# 1 Optimized parameters for the Fe-Co BOP

Table 1: Optimized parameters for the contributions to the total energy (Eq. 1) of the BOP for Fe-Co: number of valence electrons $N$ and bond integrals $dd\sigma$, $dd\pi$ and $dd\delta$ (Eq. 5) of the bond energy $E_\text{bond}$, embedding energy $E_\text{emb}$ (Eqs. 6 and 7), magnetic energy $E_\text{mag}$ (Eq. 8), pair repulsion $E_\text{rep-pair}$ (Eq. 5), core repulsion $E_\text{rep-core}$ (Eq. 9), and Yukawa repulsion $E_\text{Yuk}$ (Eqs. 10-12), as well as cutoff parameters (Eq. 13) for bond integrals ($r_{c,1}$, $d_{c,1}$) and embedding/repulsion ($r_{c,2}$, $d_{c,2}$).

|  | Parameter | Fe | Co | Fe-Co |
|---|---|---|---|---|
| $E_\text{bond}$ | $N$ | 6.98 | 8.20 | - |
| $dd\sigma$ | $c_1$ | -20.10578 | -3.37800 | -17.91242 |
|  | $\lambda_1$ | 1.68256 | 0.60345 | 1.64078 |
|  | $n_1$ | 0.83602 | 1.34934 | 0.86899 |
|  | $c_2$ | -3.09201 | -3.57946 | -2.61422 |
|  | $\lambda_2$ | 0.38260 | 0.27041 | 0.45345 |
|  | $n_2$ | 2.44521 | 4.88514 | 2.58602 |
|  | $c_3$ | 3.20957 | 3.20957 | 3.20957 |
|  | $\lambda_3$ | 0.29985 | 0.29985 | 0.29985 |
|  | $n_3$ | 6.02389 | 6.02389 | 6.02389 |
| $dd\pi$ | $c_1$ | 29.59495 | 7.71828 | 19.12989 |
|  | $\lambda_1$ | 1.47219 | 3.68495 | 1.95025 |
|  | $n_1$ | 1.27945 | -0.002336 | 0.71355 |
|  | $c_2$ | 13.96316 | 0.69810 | 14.48012 |
|  | $\lambda_2$ | 0.94631 | 0.004198 | 0.96844 |
|  | $n_2$ | 1.76772 | 5.85860 | 1.75942 |
| $dd\delta$ | $c_1$ | -10.14975 | -2.10471 | -12.45204 |
|  | $\lambda_1$ | 1.90305 | 3.85742 | 1.77809 |
|  | $n_1$ | 1.19157 | 0.051463 | 1.10349 |
|  | $c_2$ | -42.05214 | -31.39277 | -55.54683 |
|  | $\lambda_2$ | 2.58244 | 1.76031 | 2.56374 |
|  | $n_2$ | 1.07279 | 1.36317 | 1.01303 |
| $E_\text{emb}$ | $p_1$ | 4.36232 | 2.27955 | 2.78167 |
|  | $p_2$ | 0.26944 | 0.17772 | 0.20359 |
|  | $p_3$ | 0.03008 | 0.02994 | 0.03002 |
| $E_\text{mag}$ | $I$ | 0.751 | 0.841 | - |
| $E_\text{rep-pair}$ | $c$ | 155.20896 | 99.96313 | 168.93823 |
|  | $\lambda$ | 1.28421 | 1.26262 | 1.38161 |
|  | $n$ | 1.60531 | 1.68643 | 1.59143 |
| $E_\text{rep-core}$ | $p_1$ | 1.5 | 1.5 | 1.5 |
|  | $p_2$ | 100 | 100 | 100 |
| $E_\text{Yuk}$ | $B$ | 1.24913 | - | - |
|  | $r_\text{core}$ | 1.06402 | - | - |
|  | $\lambda_0$ | 1.99623 | - | - |
|  | $C$ | 113.85788 | - | - |
|  | $v$ | 1.78242 | - | - |
|  | $m$ | 1.18487 | - | - |
| cutoff | $r_{c,1}$ | 4.1 | 4.1 | 4.1 |
|  | $d_{c,1}$ | 1.0 | 1.0 | 1.0 |
|  | $r_{c,2}$ | 6.0 | 6.0 | 6.0 |
|  | $d_{c,2}$ | 0.5 | 0.5 | 0.5 |

## 2 Validation of the BOP models for pure Fe and Co

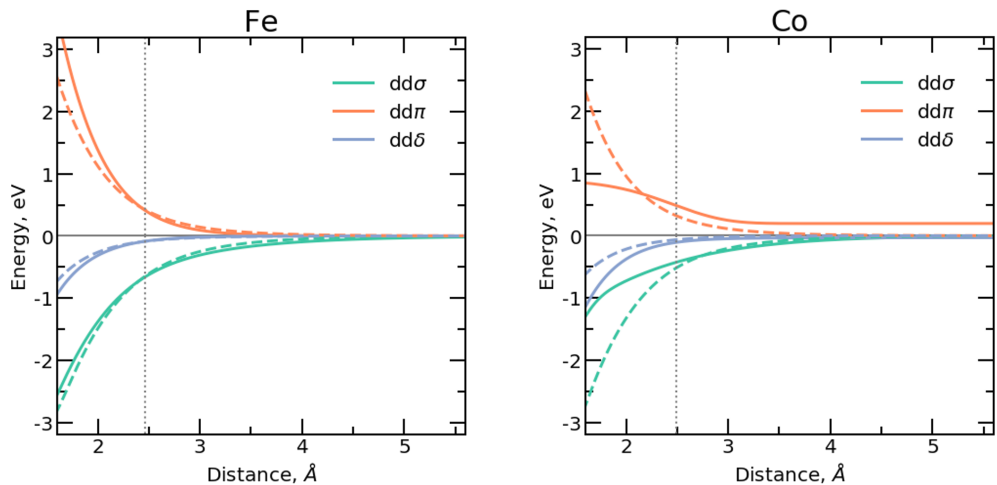

Figure 1: Bond integrals for Co-Co and Fe-Fe interactions as obtained by downfolding DFT eigenstates for a dimer [1] (dashed lines) and after optimization for bulk properties (solid lines). The first nearest neighbor distances in hcp Co and bcc Fe are indicated as dotted grey lines.

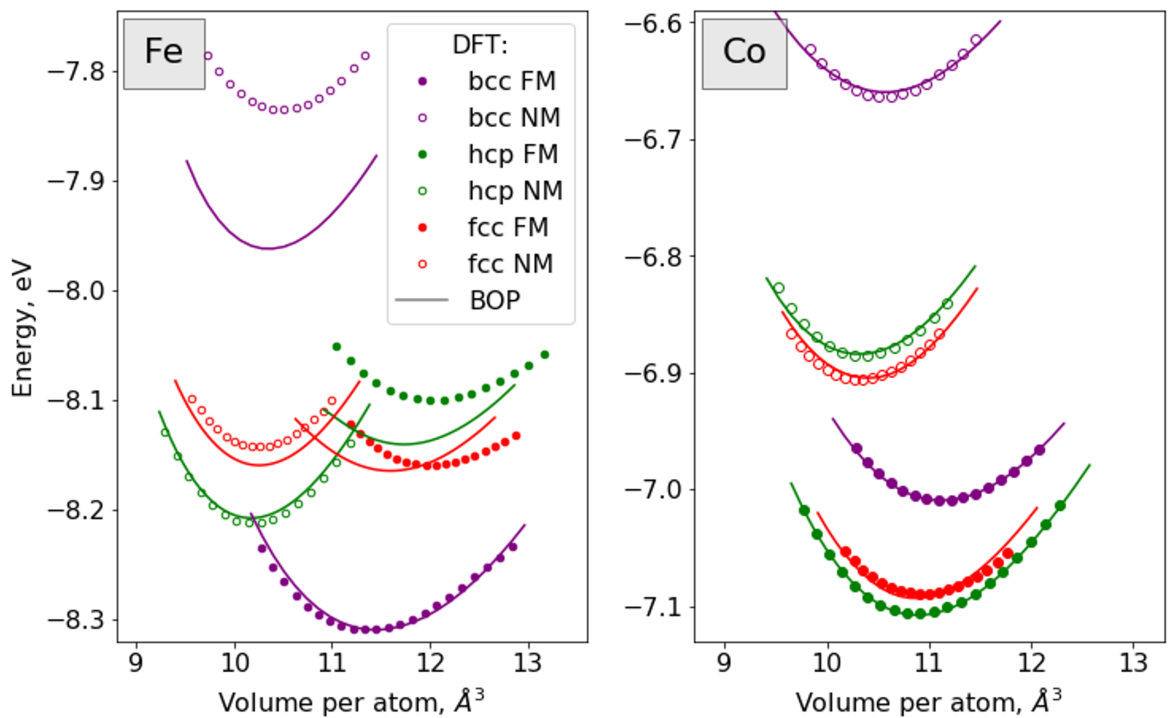

Figure 2: Comparison of energy-volume curves of Fe and Co BOP (lines) and DFT reference data points we used in the parameterization (circles).

Table 2: Volume per atom ($V$), Bulk modulus ($B$), elastic constants ($C_{11}$, $C_{12}$, $C_{44}$, plus $C_{13}$, $C_{33}$ and $C_{66}$ for hcp Co), and vacancy formation energies $E_f^v$, in bcc Fe and hcp Co obtained with BOP and DFT ($a$: [2], $b$: [3]).

|         | BOP   | DFT          |
|---------|-------|--------------|
| bcc Fe  |       |              |
| $V$     | 11.45 | $11.45^a$    |
| $B$     | 185   | $182^a$      |
| $C_{11}$| 269   | $247^a$      |
| $C_{12}$| 147   | $150^a$      |
| $C_{44}$| 90    | $97^a$       |
| $E_f^v$ | 2.15  | $2.07$-$2.22^b$ |
| hcp Co  |       |              |
| $V$     | 10.89 | $10.92^a$    |
| $B$     | 214   | $212^a$      |
| $C_{11}$| 371   | $358^a$      |
| $C_{12}$| 165   | $165^a$      |
| $C_{13}$| 130   | $114^a$      |
| $C_{33}$| 344   | $409^a$      |
| $C_{44}$| 69    | $95^a$       |
| $C_{66}$| 103   | $96^a$       |
| $E_f^v$ | 2.08  | $1.90$-$2.32^b$ |

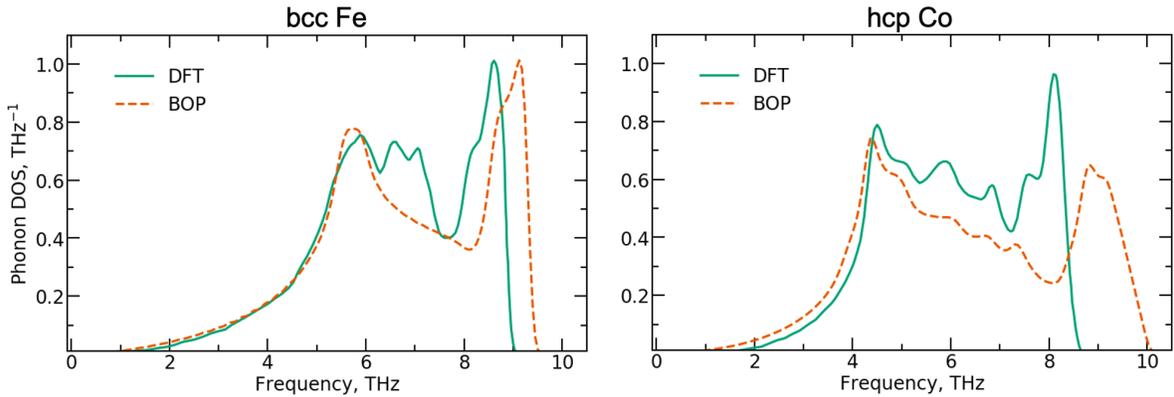

Figure 3: Phonon density of states for bcc Fe and hcp Co computed with BOP in comparison with DFT data from Ref. [4] for bcc Fe and from Ref. [5] for hcp Co.

4

# 3 Electronic DOS computed using number of moments higher than nine

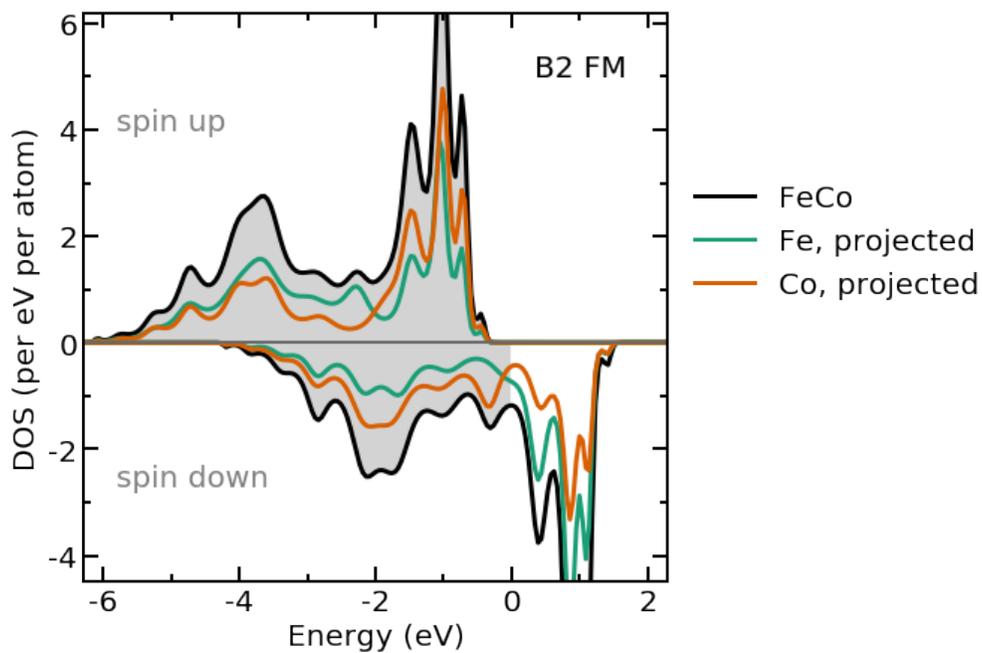

Figure 4: Electronic density of states (DOS) of ferromagnetic (FM) B2 FeCo obtained with BOP using 25 moments. The Fermi energy is taken as zero.

# 4  Point defects formation energies of FM B2 FeCo with BOP

Calculation of the formation energies of defects in ordered alloys is non-trivial compared to the pure elements due to the requirement to preserve stoichiometry. Meyer and Fähnle [6] derived an analytic expression (Eq. A7 in Ref. [6]) for calculating the formation energy of antistructure defects and vacancies for antistructure-type materials that we used to compare our Fe-Co BOP with DFT (Fig. 8 in the paper). They defined antistructure-type material as those where antistructure defects have an energy of formation significantly lower than vacancies, thus, making them a dominating defect.

To validate that B2 FeCo in our BOP model is antistructure-type, we provide an alternative calculation of formation energies at 0 K that does not rely on prior knowledge of the type of material (antistructure-type, vacancy-type or triple defect) based on a formalism suggested by Zhang and Northrup [7]. According to this formalism, the formation energy of a defect ($\Omega$) is evaluated as follows

$$\Omega = E_\mathrm{D} - n_\mathrm{Fe}\mu_\mathrm{Fe} - n_\mathrm{Co}\mu_\mathrm{Co}, \tag{1}$$

where $E_\mathrm{D}$ is the total binding energy of the block containing a defect, $n_\mathrm{Fe}$, and $n_\mathrm{Co}$ are a number of Fe and Co atoms in this block, $\mu_\mathrm{Fe}$ and $\mu_\mathrm{Co}$ are the Fe and Co chemical potentials in a defect-free B2. We assume that chemical potentials are equal to the binding energy per atom of the corresponding species in a defect-free B2 [8]. We used Formula 1 for vacancies and antisites and did all calculations with 4x2x2 supercells.

Figure 5 shows the resulting formation energies of vacancies and antistructure defects we computed with our Fe-Co BOP using Eq. 1. Formation energies of antistructure defects are significantly lower than vacancies for both Fe and Co, thus allowing B2 FeCo to be considered antistructure-type in BOP calculations. It is not a problem that Fe antistructure defect has negative formation energy because Fe and Co antisites are usually created simultaneously [6]. Thus total formation energy of this complex defect is positive.

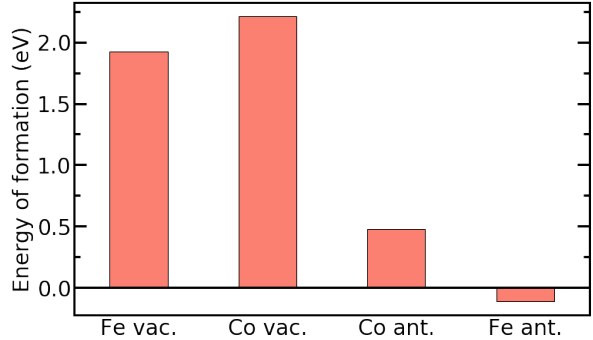

Figure 5: Energy of formation of Co and Fe vacancies and antisites in FM B2 FeCo computed with BOP.


\fi

%\end{document}

\end{document}